\documentclass[prd, singlecolumn, nofootinbib, floatfix]{revtex4-1}
\pdfoutput=1

\usepackage{hyperref}
\usepackage{amsmath}
\usepackage{graphicx}
\usepackage{dcolumn}
\usepackage{bm}
\usepackage{epsfig}
\usepackage{amssymb,latexsym,mathrsfs}
\usepackage{graphicx}
\usepackage{color}

\begin{document}

\title[Binary Black Hole Information Loss Paradox \& Future Prospects]{Binary Black Hole Information Loss Paradox \& Future Prospects}

\author{Ayan Mitra}
\email{ayan.mitra@nu.edu.kz}
\affiliation{Energetic Cosmos Laboratory,Nazarbayev University, Nur-Sultan 010000, Kazakhstan}

\author{Pritam Chattopadhyay}
\email{pritam.cphys@gmail.com}
\affiliation{Cryptology and Security Research Unit, R.C. Bose Center for Cryptology and Security,\\
Indian Statistical Institute, Kolkata 700108, India}

\author{Goutam Paul}
\email{goutam.paul@isical.ac.in}
\affiliation{Cryptology and Security Research Unit, R.C. Bose Center for Cryptology and Security,\\
Indian Statistical Institute, Kolkata 700108, India}

\author{Vasilios Zarikas$^{1,2}$}
\email{vasileios.zarikas@nu.edu.kz}
\affiliation{$^{1}$School of Engineering, Nazarbayev University Nur-Sultan 010000, Kazakhstan\\
$^{2}$University of Thessaly, Lamia 38221, Greece}


\begin{abstract}
Various techniques to tackle the black hole information paradox have been proposed. A new way out to tackle the paradox is via the use of a pseudo-density operator. This approach has successfully dealt with the problem with a two-qubit entangle system for a single black hole. In this paper, we present the interaction with a binary black hole system by using an arrangement of the three-qubit system of Greenberger–Horne–Zeilinger (GHZ) state. We show that our results are in excellent agreement with the theoretical value. We have also studied the interaction between the two black holes by considering the correlation between the qubits in the binary black hole system. The results depict a complete agreement with the proposed model. In addition to the verification, we also propose how modern detection of gravitational waves can be used on our optical setup as an input source, thus bridging the gap with the gravitational wave's observational resources in terms of studying black hole properties with respect to quantum information and entanglement.
\end{abstract}

\maketitle

\section{Introduction} \label{sec1}
At the turn of the twentieth century,  Einstein formulated the general theory of relativity (GR)~\citep{ein}. With its development, our basic understanding of the fabric of the Universe  (space-time and it's geometry) became mathematically more clear. With time, one of the strongest predictions of GR became the existence of black holes. The theory of GR is fundamentally based on the Einstein equations. It's a set of ten coupled nonlinear partial differential equations (PDE) with four independent parameters~\citep{carol,misner}.
\begin{equation}\label{a}
     G_{ab} = R_{ab} - \tfrac{1}{2} R g_{ab}=\frac{8{\pi}G T_{ab}}{c^4} + \Lambda g_{ab},
\end{equation}
where $R_{ab}$ is the Ricci curvature tensor, $R$ is the Ricci scalar, $g_{ab}$ is the metric tensor, $G$ is Newton's gravitational constant, $T_{ab}$ is the stress-energy tensor, $c$ the usual speed of light, and $\Lambda$ the cosmological constant. \footnote{Inclusion of $\Lambda$ in the Einstein equation takes into consideration of the background cosmology for a Friedmann–Lemaître–Robertson–Walker (FLRW) model. Although we don't need this term for our analysis here, but from the point of view of gravitational wave propagation, the evolution of background cosmology is governed by an FLRW universe, and hence for completeness, we presented the Einstein equation with the cosmological constant term.} Exact solutions to this set of  PDEs can describe black holes (among other things)  with different physical properties (static : Schwarzchild solution \citep{sch}, rotating : Kerr- (Newman) solution \citep{kerr}, static with electric charge : Reissner-Nordstr{\"o}m solution  (\citep{re1,re2,frolov,eric,carol,misner}) etc. However, Stephen Hawking showed \citep{hawking} that any given black hole following the principles of quantum field theory, naturally emits thermal radiations inversely proportional to its mass ($M$), with a given temperature ($T_H$) of 
\begin{equation}\label{b}
k_bT_{H}=\frac{{\hbar}}{2{\pi}\lambda_k},
\end{equation}
where $\lambda_k=2r_s/c$ is the characteristic time (in case of rotating black holes, there is an additional dependence on the angular momentum), $r_s(=\frac{2GM}{c^2})$ is the Schwarzchild radius, while $k_b$ and $\hbar$ are the usual Boltzmann constant and the reduced Planck constant.\footnote{ Hawking temperature of the black hole can be approximated from the values of the constant as $T_H\simeq10^{-7}K$}.

This is the Hawking radiation. It arises from the pair production of particles from quantum fluctuations from the horizon of the black hole. One of these particles (one with positive energy and outside the event horizon) leaves as radiation from the black hole to infinity and the other stays trapped within the black hole. As a result of the radiation, it is suggested that the black hole in the process loses mass (and hence the surface area) through the outgoing particles and hence evaporates with time. This is called as the evaporation of a black hole. Observationally it is very difficult to detect Hawking radiation as it's temperature is many orders less in comparison to the Cosmic Microwave Background (CMB) temperature $T\sim3K$, which overwhelms it (it is the reason why in last five decades of dedicated study we have not been able to still detect any such signatures of black holes).  This process, however, has some deeper consequences. For one it violates the classical Hawking area theorem~\citep{markov1} (black hole evaporation is a quantum effect) and other, an evaporating black hole, with losing mass, means that the black hole's lifetime is limited and beyond that period it potentially loses all the information that was inside it. This creates a direct violation of the quantum information conservation\footnote{Both in classical and quantum domain information conservation is fundamental, in classical physics this is governed by the Liouville's theorem of the conservation of the phase space volume \citep{susskind1}, in the quantum domain, this is preserved via the unitarity of the $S$-matrix.}. Quantum information which is quantified via the von-Neumann entropy~\citep{nc}, similar to classical physics maintains the conservation principle, that the information in a closed isolated system will be conserved~\citep{pati,clone,conserve,conserve2,conserve3}. It is intuitive to show that Hawking radiation generating from an initial pure state black hole, with the evolution of time, would end up with mixed states as remnants, thus violating the unitary evolution principle of the quantum mechanics and hence information lost during the process~\citep{ susskind00}.   If the Hawking radiation were somehow able to carry an imprint of the quantum information~\citep{clone0} from within the horizon in its flight away from the black hole to infinity, it would still give rise to new incongruency by violating the no-cloning theorem~\citep{clone1, susskind00}.

Many theories have surfaced to address this tension. One of them is the black hole complementarity principle~\citep{susskind2}, which tries to fix this problem by suggesting that the occurrence of in-falling events are temporally relative based on the observer frame, hence non-simultaneous and so unverifiable. Other theories include the holographic principle~\citep{sus2}, which states that the maximum number of states (degrees of freedom) in a confined volume is proportional to its surface area.  Recently, in their paper~\citep{monogamy}, they have proposed a new approach to tackle this problem while not disturbing the existing framework of the black hole information paradox, of the violation of monogamy principle and the black hole evaporation process occurring simultaneously. Instead, they applied a pseudo-density operator (PDO) to account for temporal and spatial entanglement between maximally entangled particles inside and outside of the black hole event horizon. With the use of the state tomography process, they simulated the scenario and successfully produced the pseudo-random operators for the model and gave measurements which were in excellent agreement with the theoretical state's value. 

In this paper, we will present a work, based on similar principles, where we will apply this formalism in a binary black hole system and show it can be successfully analyzed with a three-qubit system for binary black hole system and measurements of this generates pseudo-random state operators which are in excellent agreement with the theoretical values. We present an experimental setup for our model and perform quantum optical simulation via the quantum state tomographic process\footnote{It is not possible to estimate a quantum state from a single experimental run, due to no-clone theorem. As a result, it is necessary to reconstruct the state multiple times and do the measurements number of times on a different basis. Basic state tomography involves the estimation of the expectation values of all the operators (we parameterize any given quantum states of a system with respect to a set of operators), and if one can reconstruct all the operators then the experiment is said to be tomographically complete.}~\citep{tomo}. So far we have been considering the situation where there is no correlation between the two qubits of the binary black hole system.  Now, we have considered a situation that there exists a correlation between the qubits of the binary black-hole. This can be described by using the pseudo-density operator formalism by considering the interaction between the qubits of the binary black hole with the particle above the event horizon. Interestingly the results show an excellent agreement with our proposed theoretical proposal.

The paper is structured as follows.  In section \ref{sec2}, we describe the pseudo-density operator formalism which is more general than a known density operator. This formalism guides to explain the binary black evaporation theory. We have devoted section \ref{sec3} for the analysis of this formalism by simulation of the optical setup. We encounter that the formalism is able to explain the binary black hole evaporation, which violates the monogamy principle of entanglement theory. Even the proposed model is able to explain the correlation of the qubits within the black hole when the interaction between the two black hole system is considered. In section \ref{sec4}, we put forward the recent results from the literature of how modern gravitational wave detection could possibly be used to extract information about black hole radiations, and then suggest how our experiment can be possibly linked with future high precision gravitational waves detection programs. We conclude the paper in section \ref{sec5} with some discussion that has been analyzed in this paper.

\begin{figure}
	\includegraphics[width=\columnwidth]{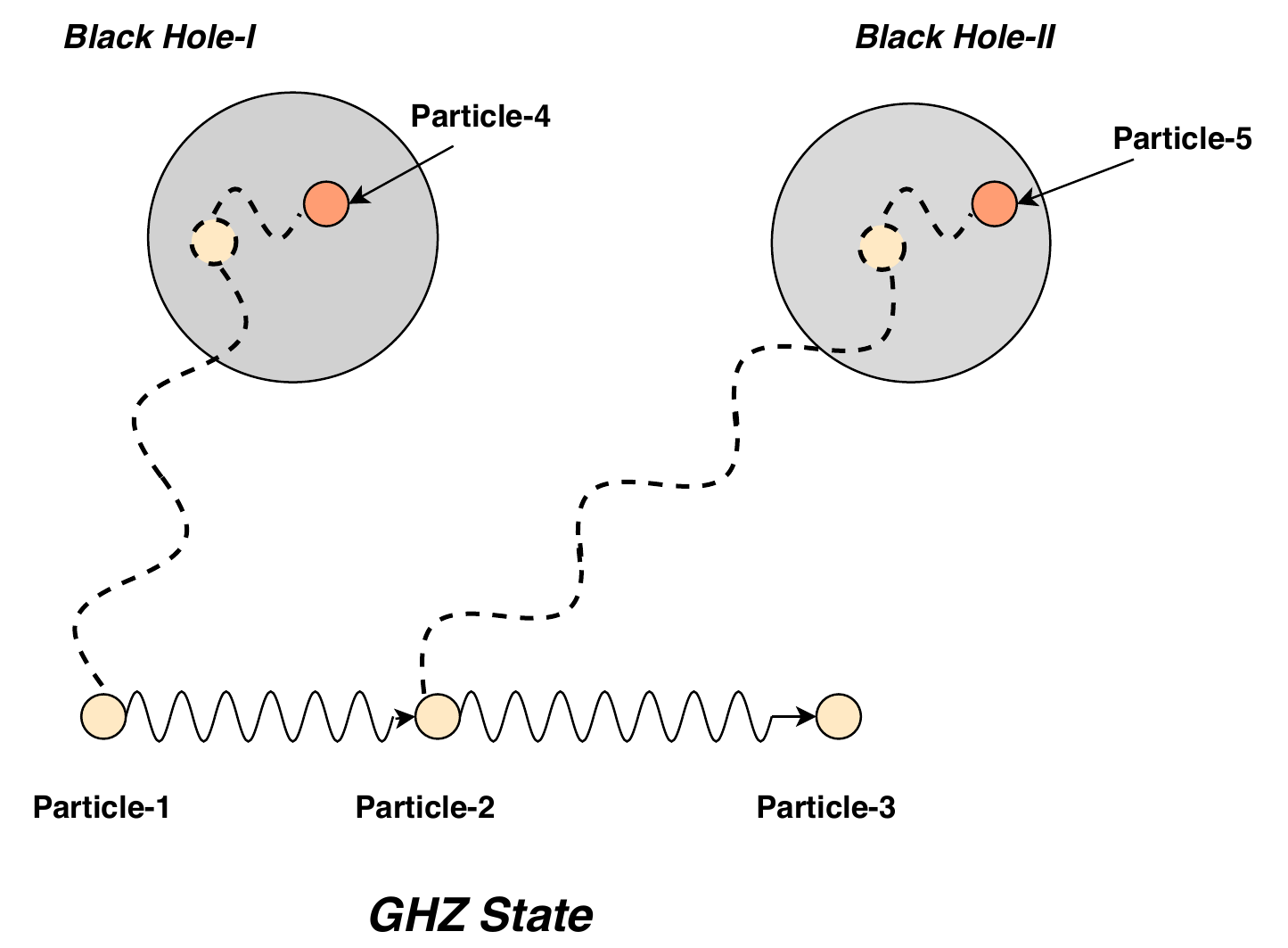}
    \caption{This is the schematic representation of the process of the black hole evaporation for a binary system from a pseudo-density operator framework.}
    \label{f2}
\end{figure}

\begin{figure}
	\includegraphics[width=\columnwidth]{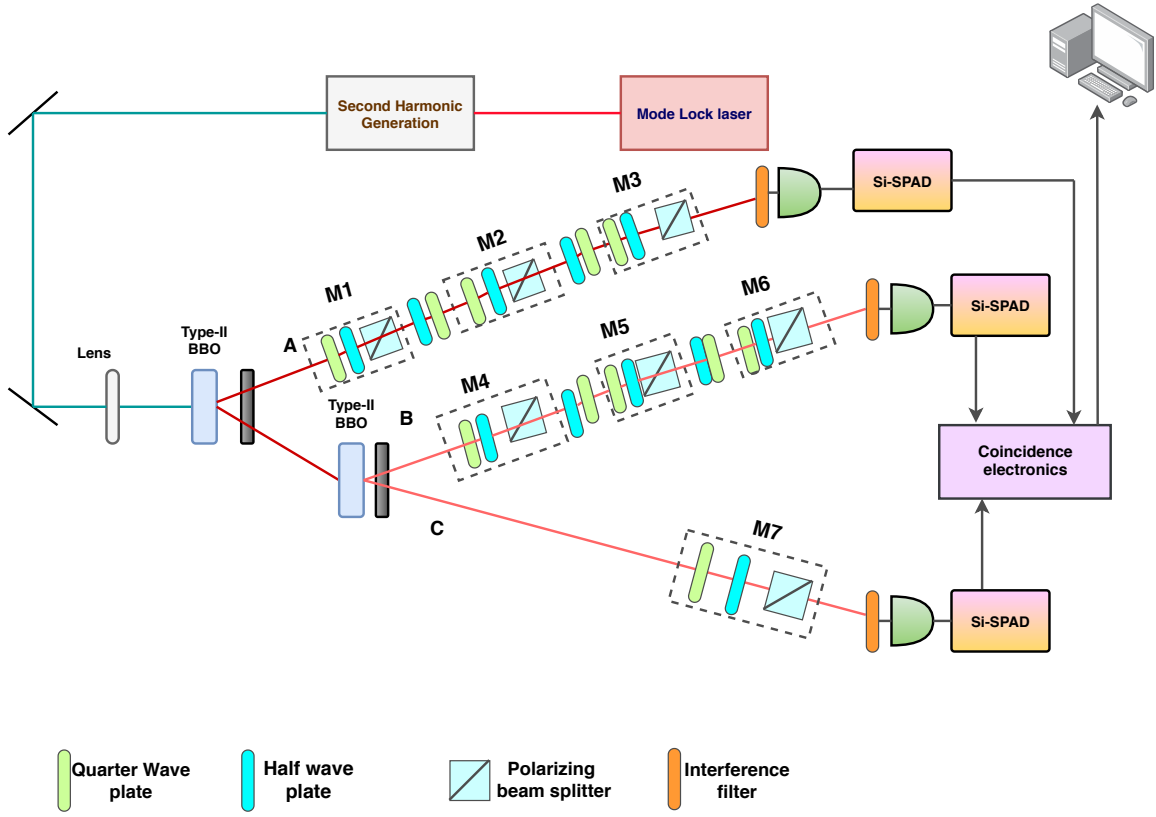}
    \caption{Experimental setup of the process. Here a GHZ state is generated by using two sets of $\beta-$barium borate (BBO) type-II crystal. Three sets of measurements are considered on photon A, B, where the measurements are considered for three different times ($t_1$,  $t_2$ and $t_3$ respectively) and a single measurement for the photon C.}
    \label{f3}
\end{figure}

\section{Background}\label{sec2}
It is a known fact that the Schwarzschild metric describes the space-time continuum in the presence of the black hole. The Schwarzchild metric in a three dimensional ($r,\theta,\phi$) spherical coordinate system is given as,
\begin{equation}
    ds^2=-Kdt^2 + \frac{1}{K}dr^2+d\Omega^2,
\end{equation}
where $K=\left(1-\frac{2M}{r}\right)$ and $d\Omega^2=r^2d\theta^2+r\sin{\theta}^2d\phi^2$, where $r\ge r_s$ (described in Eq. \eqref{b}) and $M$ is the corresponding spherical mass at radius $r$.  A particle crossing the horizon is equivalent to swapping of the signature of the metric, i.e., the spatial and the temporal components~\citep{misner}. Now in the quantum realm, if one considers a quantum phase factor, then the change in the spatial and the temporal is simply conveyed by the conjugate of the defined phase factor.  So, the transpose operation of the density matrix can describe the effect of the in-falling quantum system. The transpose operation so defined is a positive operation but defies to represent a completely positive operation, which indicates that if one performs a transpose operation on one of the three-party entanglement system, the state of the system may not turn out being a valid density matrix. For this, pseudo density operators (PDO) are used to explain this phenomenon~\citep{fitz},  which can accommodate non-positive operations like Hermitian transformations as well. We are going to exploit this fact to neutralize the violation of the monogamy principle of the entanglement theory during the evaporation of the black hole.

In this paper, we consider a maximally entangled Greenberger–Horne–Zeilinger (GHZ) state (three-qubit system) and a binary black hole system with pure states. We name the three particles maximally entangled as particle $1$ and so on. We consider that two particles from this system fall in the binary black hole system, as shown in Fig. (\ref{f2}). Particle $1$ falls in black hole $1$ and particle $2$ in black hole $2$. Once inside, the particles will entangle with particles from inside the black hole environment, we name them particle $4$ and $5$ in the two black holes successively.  We implement this setup as per the optical setup shown in Fig. (\ref{f3}) and then, we do the tomographic reconstruction of the state to analyze the black hole evaporation from an information theory standpoint. The simulation returns a pseudo-density matrix which can then be compared to our true value pseudo-density operator ($\rho_{true}$, which depicts the theoretical expectation of the state) via a distance measure between the two and give a figure of merit on the comparison of measurement values between the particle entangled inside the black hole and that outside. We present this comparison in terms of a fidelity score. We show that the fidelity score is sensitive to the method of estimation that are used in our analysis. We have used three different methods: maximum likelihood, and two variants of linear inversion techniques to do the state tomography, yet our overall fidelity score is excellent, inferring that it is possible to do the measurement of the particle that is inside the black hole via the measurement of the particle that is outside.

A density matrix bestows the probability distribution of the pure states, i.e., $\rho = \sum_j a_j |\phi_j\rangle \langle \phi_j|$, where $a_j$ describes the probability of the pure state $|\phi_j \rangle$. The expectation value of a Pauli matrix is defined as $\langle a \rangle = tr(a \rho)$. So, we can describe an alternative approach to formulate the density matrix in terms of the Pauli operator. So, for an $n$-qubit system, the general density operator in terms of the Pauli operators is defined as 
\begin{equation} \label{c}
    \rho_n = \frac{1}{2^n} \sum_{\alpha_1 = 0}^3 \cdots \sum_{\alpha_n=0}^3 \langle \bigotimes_{\beta=1}^n \sigma_{\alpha_{\beta}} \rangle \bigotimes_{\beta=1}^n \sigma_{\alpha_{\beta}},
\end{equation}
where $\sigma_0 = {\mathbb I}$, $\sigma_1 = X$, $\sigma_2 = Y$, $\sigma_3 = Z$. Whereas the PDOs generalises these operators and contains the statistics of the time domain. A general form of the PDO for a $n$-qubit is described as:
\begin{equation} \label{d}
    P_n = \frac{1}{2^n} \sum_{\alpha_1 = 0}^3 \cdots \sum_{\alpha_n=0}^3 \langle  \{\sigma_{\alpha_{\beta}}\}_{\beta=1}^n \rangle \bigotimes_{\beta=1}^n \sigma_{\alpha_{\beta}}.
\end{equation}
 If one consider a set of event $\{E_1, E_2, \cdots, E_m\}$, for each event $E_\beta$ we can have a single qubit Pauli measurement operator $\sigma_{\alpha_{\beta}} \in \{ \sigma_0,\cdots, \sigma_3\}$. Now for any specific choice of Pauli measurement operator $\{\sigma_{\alpha_{\beta}}\}_{j=1}^{n}$, we consider $\langle  \{\sigma_{\alpha_{\beta}}\}_{\beta=1}^n \rangle$ as the expectation value product of the result of these measurements. This can be in space or in time. The PDOs shares many properties in common with the density matrix. All PDOs are necessarily Hermitian in nature, trace one. The main difference of the PDOs with the density matrix is that they are not necessarily positive operators, i.e., they can possess negative eigenvalues. 

We will now try to comprehend the working principle of PDOs relevant to the problem under study. Let us consider a maximally mixed state for a three-qubit system. Now, we will describe a physical process where a system of qubits is measured at two different times. The measurements are performed in the complimentary Pauli bases $X$, $Y$, and $Z$. The outcome of the measurement statistics can be expressed by an operator, the quantum density operator. This quantum density operator is the pseudo-density operator~\citep{fitz} which is described as 

\begin{equation}\label{e}
    P_{123} = \frac{1}{8} [I + X_1X_2X_3 + Y_1Y_2Y_3 + Z_1Z_2Z_3],
\end{equation}
where the subscripts indicate the index of qubits. One can obtain the reduced state of the subsystem by tracing out the subsystem whose information is not of concern. Surprisingly, one can represent the pseudo-density operators by executing a partial transpose operation over the maximally entangled basis of the respective dimension. We use this model to understand what happened to the three-particle entangled qubits when two of the qubits are falling into the binary black hole system. We use $P_{123}$ to describe the state of the system where it is considered that two of the qubits is falling into the binary black hole. This is schematically explained in Fig. (\ref{f2}).

Based on this reasoning, we will propose a PDO to model the problem under execution. Here, a three-qubit entangled state is considered, out of them, two of the particles gets further entangled with two other particles in the binary black hole system. We would explain that the black hole information problem and binary black hole system can be explained by contemplating the PDO model, which is represented by Eq.~\eqref{e}. This PDO represents a three-qubit entangled system, out of which two of the entangled particles cross the event horizon and fall into the black hole and there the particles get entangled with a qubit. This proposed method describing the correlations associated with the black-hole evaporation is in agreement with that proposed by \citep{monogamy}, and the explanation of the black hole ringdown stage boils down to the equivalent two-qubit system from the three-qubit system.

\begin{figure*}
\centering
	\includegraphics[width=\textwidth]{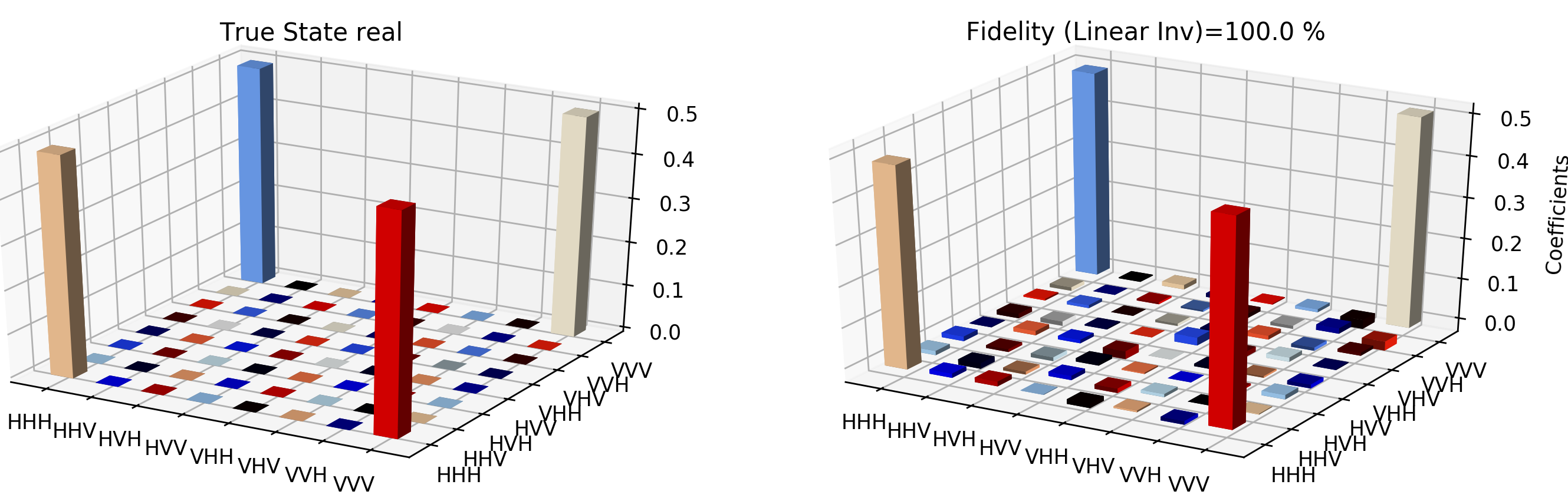}
    \caption{Tomographic reconstruction of the reduced pseudo-density operator $P_{143}$ using the linear inversion method. The real part of the theoretical expectation (depicted by the true state in the plot) and the real part of the reduced pseudo-density operator is compared here.}
    \label{f4}
\end{figure*}
\begin{figure*}
\centering
	\includegraphics[width=\textwidth]{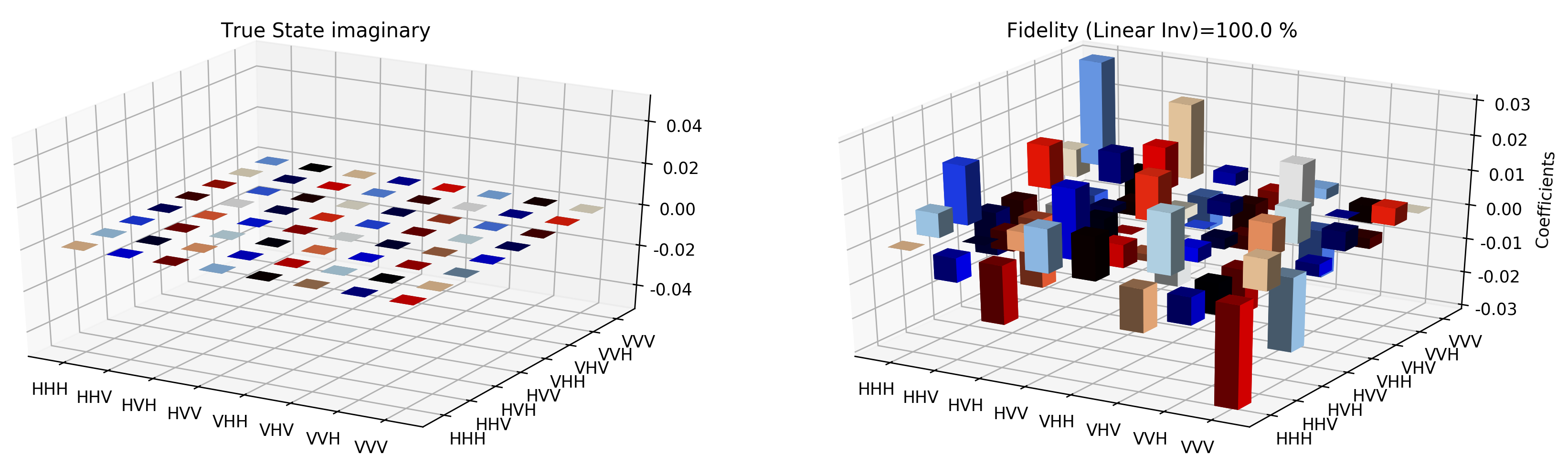}
    \caption{Tomographic reconstruction of the reduced pseudo-density operator $P_{143}$ using the linear inversion method. The imaginary part of the theoretical expectation (depicted by the true state in the plot) and the imaginary part of the reduced pseudo-density operator  is compared here.}
    \label{f5}
\end{figure*}

\begin{figure*}
\centering
	\includegraphics[width=\textwidth]{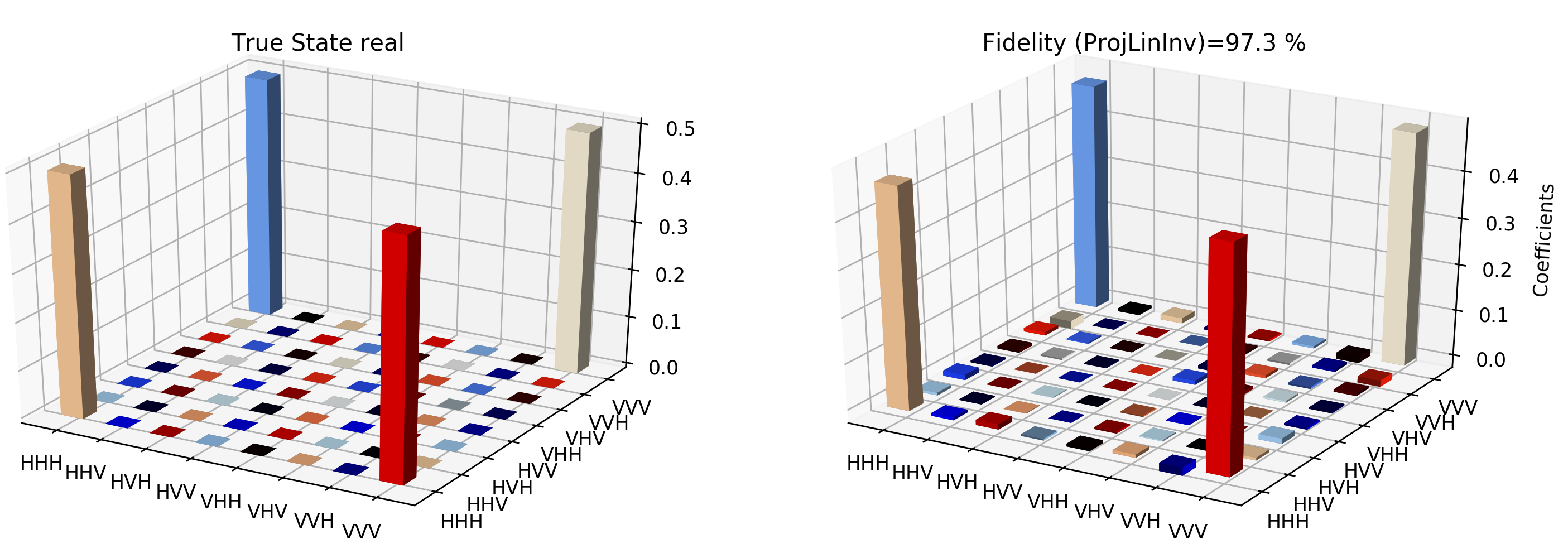}
    \caption{Similar to Fig.\ref{f4}, state tomography reconstruction of the reduced pseudo-density operator $P_{143}$ is conducted using the projected linear inversion method. The real part of the theoretical expectation (depicted by the true state in the plot) and the reduced pseudo-density operator is compared.}
    \label{f6}
\end{figure*}

\begin{figure*}
\centering
	\includegraphics[width=\textwidth]{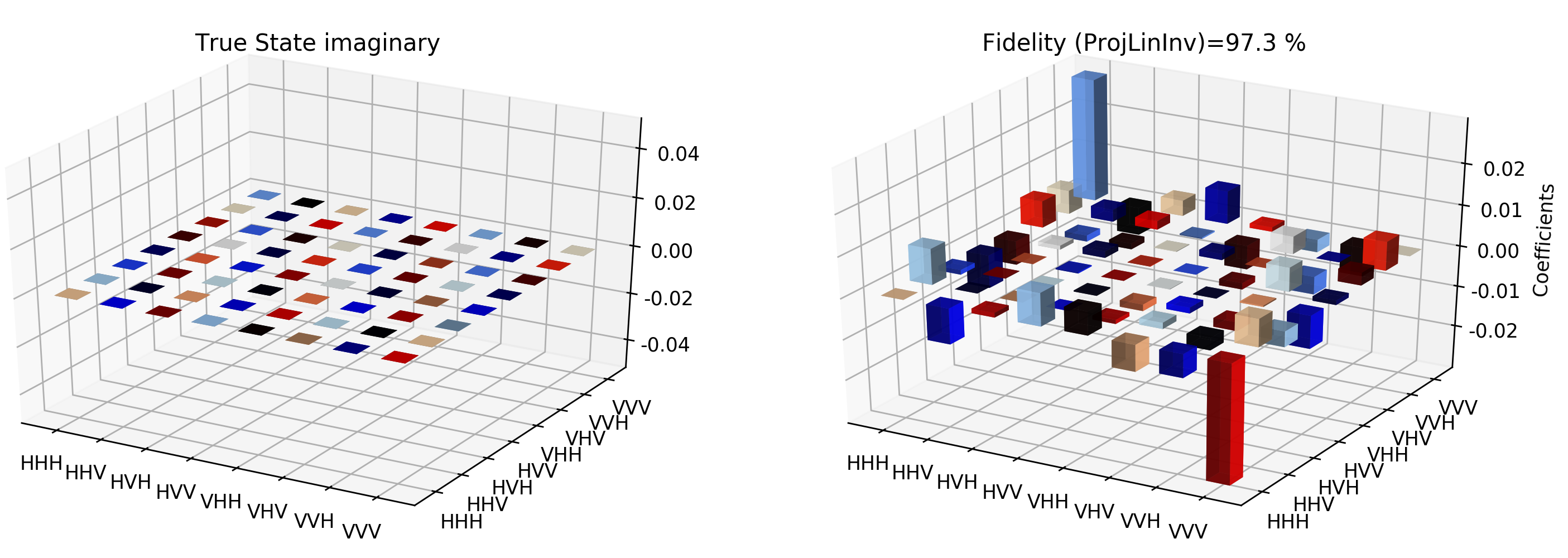}
    \caption{Similar to Fig.\ref{f5}, state tomography reconstruction of the reduced pseudo-density operator $P_{143}$ is conducted using the projected linear inversion method. The imaginary part of the theoretical expectation (depicted by the true state in the plot) and the reduced pseudo-density operator is compared.}
    \label{f7}
\end{figure*}

\begin{figure*}
\centering
	\includegraphics[width=\textwidth]{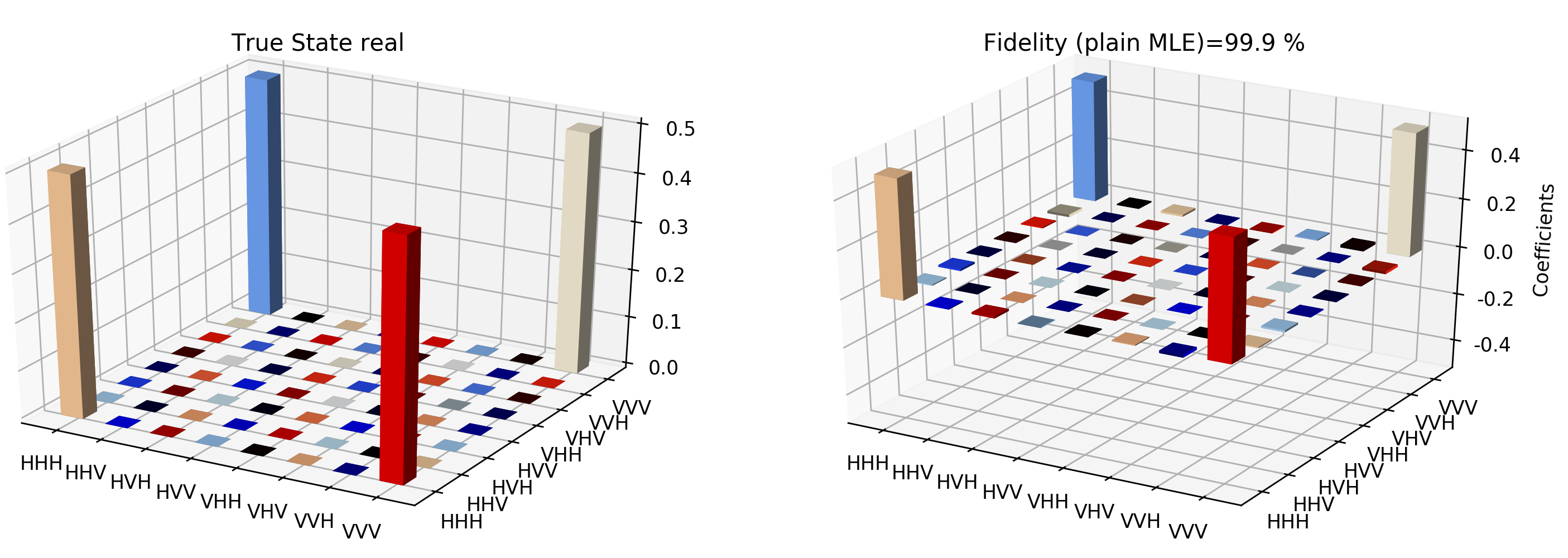}
    \caption{State tomography reconstruction of the reduced pseudo-density operator $P_{143}$ is conducted using the maximum likelihood estimation method. The real part of the theoretical expectation (depicted by the true state in the plot) and the reduced pseudo-density operator is compared.}
    \label{f8}
\end{figure*}

\begin{figure*}
\centering
\includegraphics[width=\textwidth]{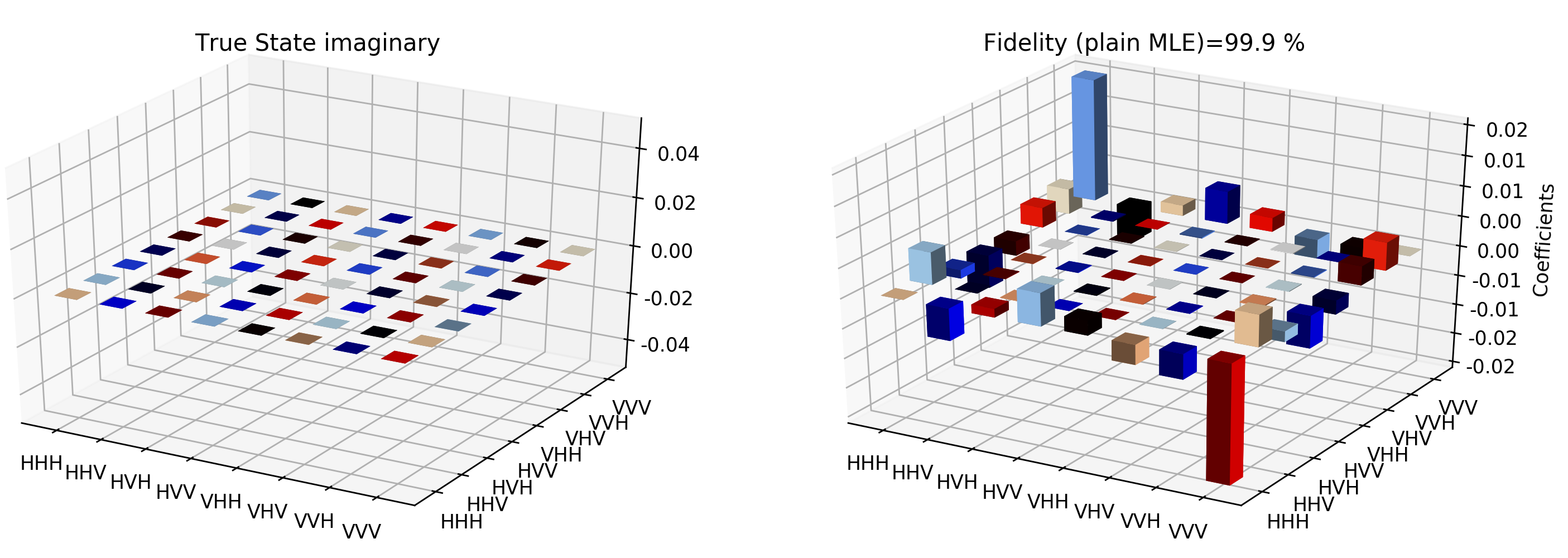}
    \caption{ State tomography reconstruction of the reduced pseudo-density operator $P_{143}$ is conducted using the maximum likelihood estimation method. The imaginary part of the theoretical expectation (depicted by the true state in the plot) and the reduced pseudo-density operator is compared.}
    \label{f9}
\end{figure*}

\begin{figure*}
\centering
	\includegraphics[width=\textwidth]{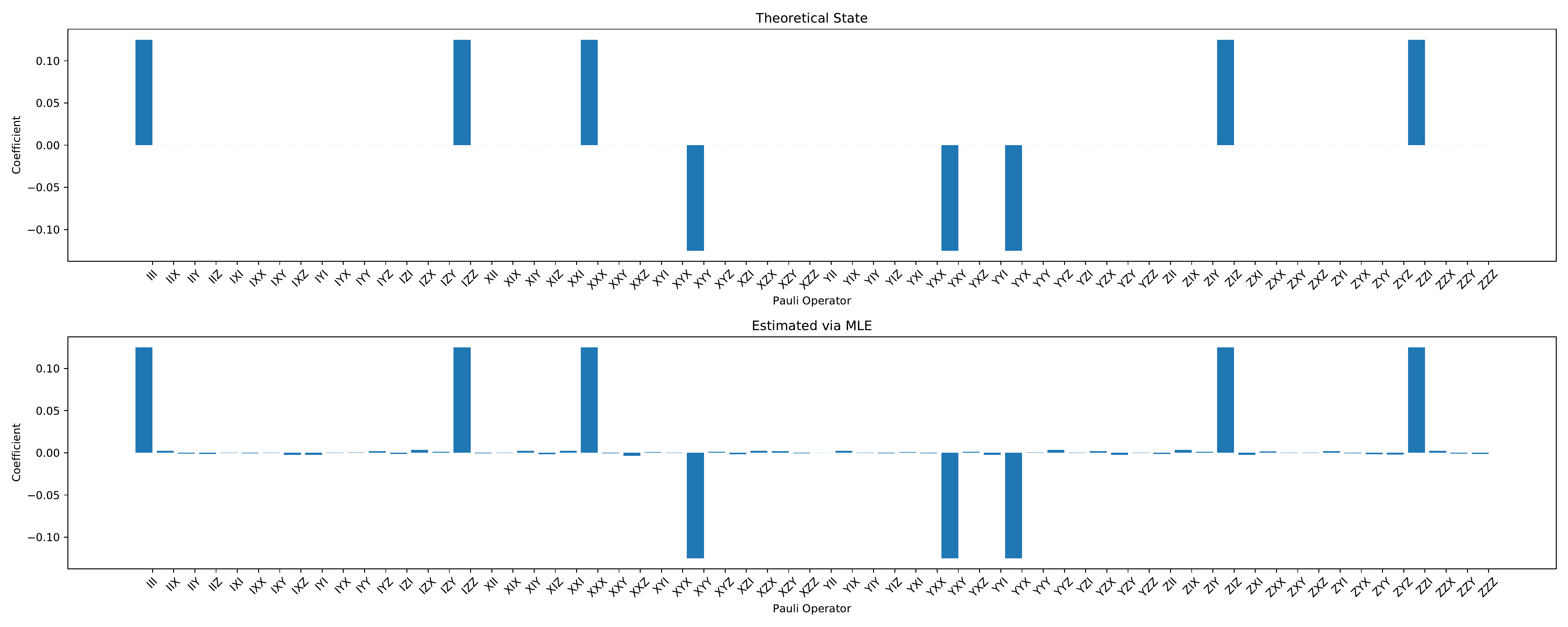}
    \caption{The comparison of state tomographic reconstruction of the pseudo-density operator $P_{143}$ and the theoretical state (depicted by the true state in the plot) after the execution of the measurement.}
    \label{f10}
\end{figure*}

\begin{figure*}
\centering
\includegraphics[width=\textwidth]{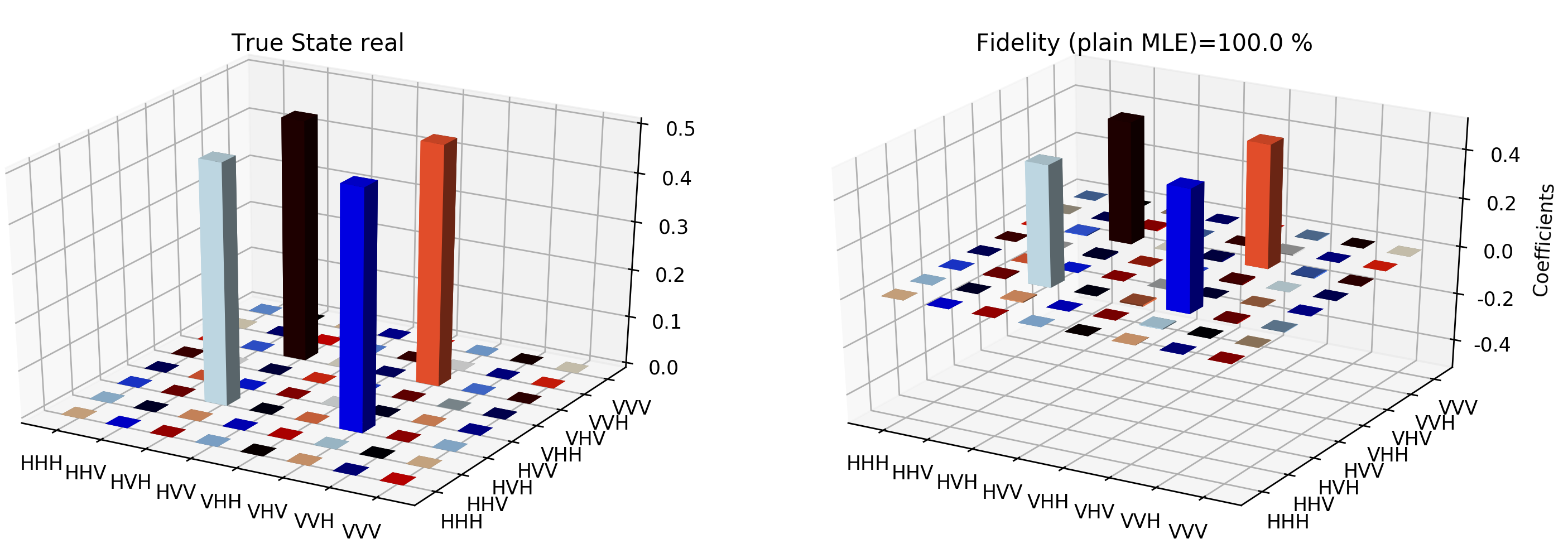}
    \caption{ State tomography reconstruction of the reduced pseudo-density operator $P_{453}$ is conducted using the maximum likelihood estimation method. The imaginary part of the theoretical expectation (depicted by the true state in the plot) and the reduced pseudo-density operator is compared.}
    \label{f11}
\end{figure*}

\begin{figure*}
\centering
\includegraphics[width=\textwidth]{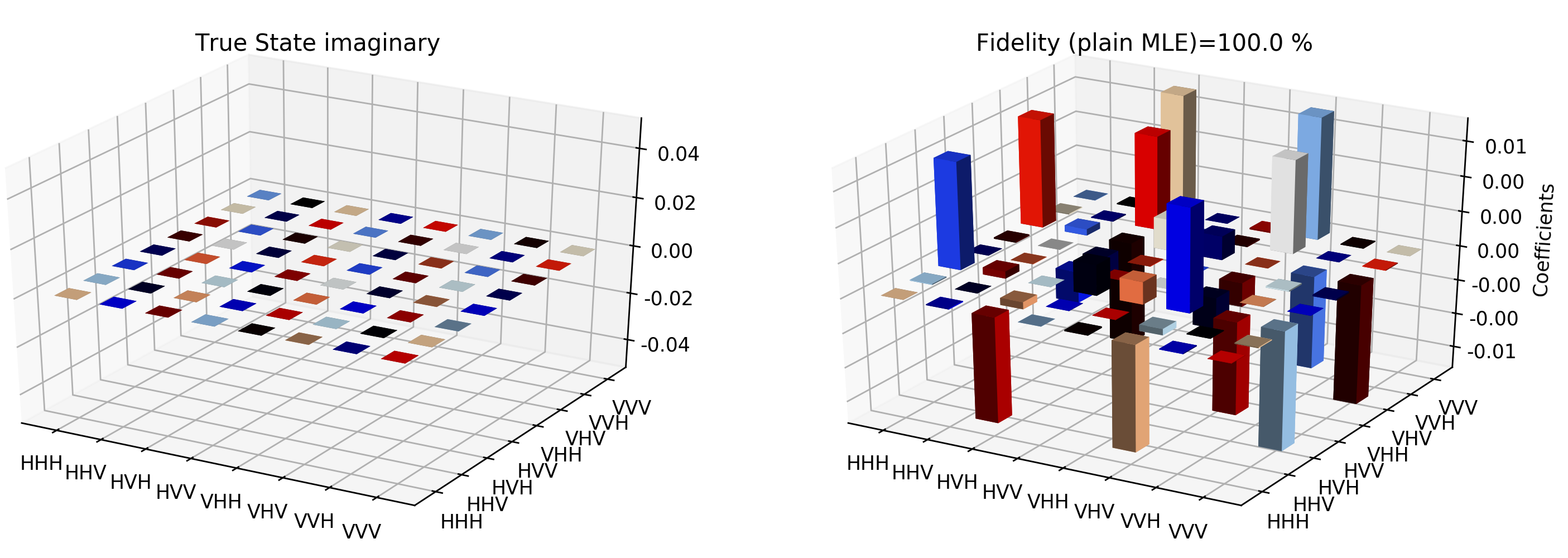}
    \caption{ State tomography reconstruction of the reduced pseudo-density operator $P_{453}$ is conducted using the maximum likelihood estimation method. The imaginary part of the theoretical expectation (depicted by the true state in the plot) and the reduced pseudo-density operator is compared.}
    \label{f12}
\end{figure*}

\section{Analysis}\label{sec3}
We are going to explain the binary black hole evaporation theory with the help of the PDO model.  We will take into account that a three-qubit entangled state is created above the event horizon. Now one of the particles of the GHZ state that is created due to the process of Hawking radiation falls into one of the black holes of the binary black hole system, and the second particle falls in the second black hole. Time like correlation is developed between them. Now when the two particles that have fallen in the black hole get entangled respectively with a qubit in the black hole, the system can be represented by a five-qubit entangled pseudo-state. The total pseudo-density operator for the system can be described as 

\begin{equation}\label{f}
    P_{12345} = \frac{1}{2^5} \Big[I + \Sigma_{123}- \Sigma_{143}-\Sigma_{413}-\Sigma_{253}-\Sigma_{523} \Big],
\end{equation}
where $\Sigma_{ijk}= X_iX_jX_k + Y_iY_jY_k +Z_iZ_jZ_k$. The Eq.~\eqref{f} is based on the framework outlined in ~\cite{monogamy}. According to the conjecture proposed in the work~\cite{monogamy}, they have considered that the time-like correlation is positive, and whereas the spatial correlation component is negative. This is based on the metric signature convention in general relativity, which typically follows the $[+,-,-,-]$ (or $[-,+,+,+]$) convention where the positive sign is for the temporal component and the remaining three negative signs are for the spatial component. Following the same convention, we have defined the pseudo density operator for our system in Eq.~\eqref{f}.  The correlation described by the pseudo density for this system does not obey the monogamy principle of entanglement theory. We will now use this PDO to explain the binary black hole system and discuss how the merger of the black hole boils down equivalent to the two-qubit system.

So far in the analysis of the binary black hole system, the correlation between the qubits of the two black hole was not taken under consideration. Here we will consider the case, where the correlation between the two qubits (interaction term) in the binary black hole systems are taken into account. The pseudo-density operator with this correlation is expressed as
\begin{equation}\label{j1}
     P_{12345} = \frac{1}{2^5} \Big[I + \Sigma_{123}- \Sigma_{143}-\Sigma_{413}-\Sigma_{253}-\Sigma_{523} -\Sigma_{453} \Big],
\end{equation}
where the term $\Sigma_{453}$ represents the correlation of the qubits of the two black hole systems. Similar to the process conducted above for the analysis of $P_{123}$, we execute the state tomographic reconstruction of the state $P_{453}$, which can be obtained from Eq.~\eqref{j1} by tracing out the information of the particle one and two (which can be depicted as $P_{453}=\frac{1}{8} (\mathbb{I} - \Sigma_{453})$).

If two-qubits systems (like A and B) are maximally correlated they cannot be correlated with a third qubit C. For this convention, there exists a trade-off between the amount of entanglement between the qubits A and B, and the same between the qubits A and C. One can express this mathematically using the Coffman-Kundu-Wootters (CKW) monogamy inequality~\citep{coffman12,oso} as

\begin{equation}\label{g}
   C_{AB}^2+ C_{AC}^2 \leq C_{A(BC)}^2,
\end{equation}
where $C_{AB}$, $C_{AC}$ represents the concurrences between A and B, and between A and C respectively, while $C_{A(BC)}$ is the concurrence between subsystems A and BC. $C_{AB}$ is defined as $C_{AB} = max \{0,\lambda_1 - \lambda_2 - \lambda_3 - \lambda_4\}$. Here the $(\lambda_i)$ represents the square root of the eigenvalues of the matrix $\rho_{ij} (\sigma_y \otimes \sigma_y) \rho_{ij}^{\star} (\sigma_y \otimes \sigma_y)$, where $\rho_{ij}^{\star}$ depicts the complex conjugate of the density matrix and $\sigma_y$ the Pauli matrix.  The monogamy inequality can also be expressed in terms of entanglement measures as
\begin{equation}\label{h}
    E(A|B)+ E(A|C) \leq E(A|BC).
\end{equation}
For $N$ qubit~\citep{Zhu1} the definition can be extended as

\begin{equation}\label{i}
    E(A|B_1)+ E(A|B_2) + \dots + E(A|B_{N-1}) \leq E(A|B_1B_2\dots B_{N-1}).
\end{equation}
Using the equation (\ref{i}), we can analyze the monogamy inequality for our system. This is violated by our pseudo operator $P_{12345}$.

The above proposed PDO describes the binary black hole evaporation which incorporates the monogamy violation principle. This is possible because PDOs can be used to describe the maximally temporal correlation as well as maximally spatial correlation.

To describe this process, we execute a quantum optical simulation of this framework. Here, we are not going to describe any experimental test results, but we will illustrate our theoretical model via qubit simulation using quantum virtual machines. Through our experiment, we first generate a three-particle entangled pair of photons ($A$, $B$, $C$). Now, after the two-particle falls into the black hole the correlation between the individual particles that have fallen and the particle that is above the event horizon is in the same maximally entangled state, which is observed by measuring the photon A and B in three different times ($t_1$, $t_2$ and $t_3$). Whereas, the correlation between the particles that have fallen inside the black hole, and has developed a spatial entanglement there, which can be comprehended by measuring the photons $A$, $B$, $C$ at the same time $t_1$. So, we reconstruct the relevant statistics of the PDO $P_{12345}$. This is established by constructing the different ensemble of the particles under study.

In the optical schematic, we have generated a GHZ state using a type-II BBO crystal~\citep{deny}. A mode lock laser has been considered for the generation of a laser beam of 808 nm wavelength. This beam is then passed through a second harmonic generator after which it gets injected into a 0.5 nm thick BBO crystal of type-II to generate a parametric down-conversion (PDC)~\citep{rubin1,deny}. After the generation of the two-photon beam, the second photon beam is again injected into a BBO crystal to produce two further beams. These generate a three-photon entangled state. The maximally entangled state is $|GHZ\rangle = \frac{1}{\sqrt{2}} \Big(|HHH\rangle + |VVV\rangle \Big)$, where $H$ and $V$ represents the horizontal and the vertical polarisation components respectively. These are generated from the interaction of the PDC cone~\cite{Genovese}.

In two of the photon paths ($A$, $B$), two sets of measurements is conducted here in cascades ($M_1$, $M_2$, $M_3$ for photon path $A$ and $M_4$, $M_5$, $M_6$ for the photon path $B$). Each of these measurement systems when unfolded, consists of a quarter-wave plate (QWP), then a half-wave plate (HWP), and a polarizing beam splitter (PBS). We have inserted a set of HWP and QWP between two measurements so that, one can compensate for the polarisation that occurred due to the previous measurement. After the measurement, the photon $A$, $B$, and $C$ are passed through the band-pass interference filter, which filters the photon beam. After the filtration process, it passed through the multi-mode optical fibers connected to silicon single-photon avalanche diodes (Si-SPADs). The output is then sent to the coincidence electronics for the analysis.

We will perform a quantum state tomography reconstruction~\citep{bog,daniel12} on branch A. In this case, we are able to extract the temporal correlation for the system which can be described as $P_{123} = \frac{1}{8}(I+ \Sigma_{123})$ and to understand the spatial correlation we have conducted a tomographical reconstruction of the reduced pseudo density state $P_{143}$ of the system. Similarly, one can develop the other reduced pseudo-density state by a similar chronology.

The state tomographic reconstruction of the state $P_{123}$ is shown in Figs. (\ref{f4}-\ref{f9}). For the analysis, we have considered three different methods to estimate and reconstruction of the state. The results so generated using these methods are in excellent agreement with the theoretical expectations as stated by the fidelity ($F$), which is the measure that evaluates the closeness of the state expressed by the density matrix to that of the original pure state $|\psi\rangle$, $F$ can have a value between $[0,1]$. For Maximum Likelihood Estimation (MLE), linear inversion and projected linear inversion the fidelity is $F= 99.9\%$, $100\%$ and $97.3\%$ respectively (shown in table (\ref{table1})). The state tomographic reconstruction of the state $P_{143}$ results similar to $P_{123}$. The simulation of the monogamy inequality of the considered pseudo-density matrix shows that it violates the monogamy principal. The detailed plots of the analysis of $P_{143}$ are not shown as they are similar in nature.

\begin{table}
    \centering
     \begin{center}
 \begin{tabular}{||c c c c||} 
 \hline
 Method & Fidelity Score  \\ [0.7ex] 
 \hline\hline
 Linear Inversion & 1.0 \\ 
 \hline
 Projected Linear Inversion & 0.973 \\
 \hline
 MLE & 0.999 \\
 \hline
\end{tabular}
\end{center}
    \caption{Table showing the fidelity score $F$ obtained from the three different methods used in the tomography used. Since $F$ can't exceed values of $0.5$ in the classical limit, it shows that there is true entanglement beyond the classical limit. Also, the deviation in the models shows that better entanglement distillation could resolve this difference in values.}
    \label{table1}
\end{table}

It is however interesting to note the fluctuations in the imaginary part of each of the plots. Although the absence of any fluctuations in the real axis compels us to believe that it is simply not background noise, originating from measurement error.  If we compare our imaginary plot results to that of the~\citep{monogamy} plots, we see clearly there are much more fluctuations in our binary black hole system set up. It is not clear to us at the moment what are the origins of these fluctuations, but definitely, it points to some perturbations on the quantum state measurements of the pseudo-random operators originating specifically from our system's set up (hence essentially a quantum phenomenon). We speculate this could be any deviations around the horizon of the black hole. In the future, we plan to verify this analysis with an optomechanical setup~\cite{Bosso,piko12} and further explore in the Planck regime for any possible deviation in the horizon of the black hole. We look forward to studying the cause of such anomaly in the imaginary axis values and exploring it further in future works along with a similar framework as presented by~\citep{echo}.

For the analysis of the interaction between the two qubits of the two black hole systems, we have considered a different basis of the GHZ state \citep{mcunha}, from which we can return to the usual form by some local operation. One can obtain the maximally entangled state by adopting the selected photons spatially which belongs to the intersection of two parametric down-conversion cones. The process properly compensates for the temporal and the phase effect~\cite{Genovese}. To measure the spatial correlation (like $P_{453}=\frac{1}{8} (\mathbb{I} - \Sigma_{453})$) we measure the correlation between $M_4$, $M_5$ and $M_7$, provided that $M_6$ performs the same polarization projection as that of $M_4$ and $M_5$. By this process, we reconstruct the reduced pseudo-density operator tomographically which actually corresponds to the spatial entangled state that is formed between the particles 4, 5, and 3 within the black hole.
The state tomographic reconstruction of the system shows a complete agreement with the proposed theoretical model for the analysis as shown in Fig. (\ref{f11}).  Similar to the state tomographic analysis of $P_{123}$, we also encounter fluctuations in the imaginary part of the plot which we can speculate as to the effect due to the interaction of the two qubits in the binary black hole system.

\begin{figure}
\centering
	\includegraphics[width=\columnwidth]{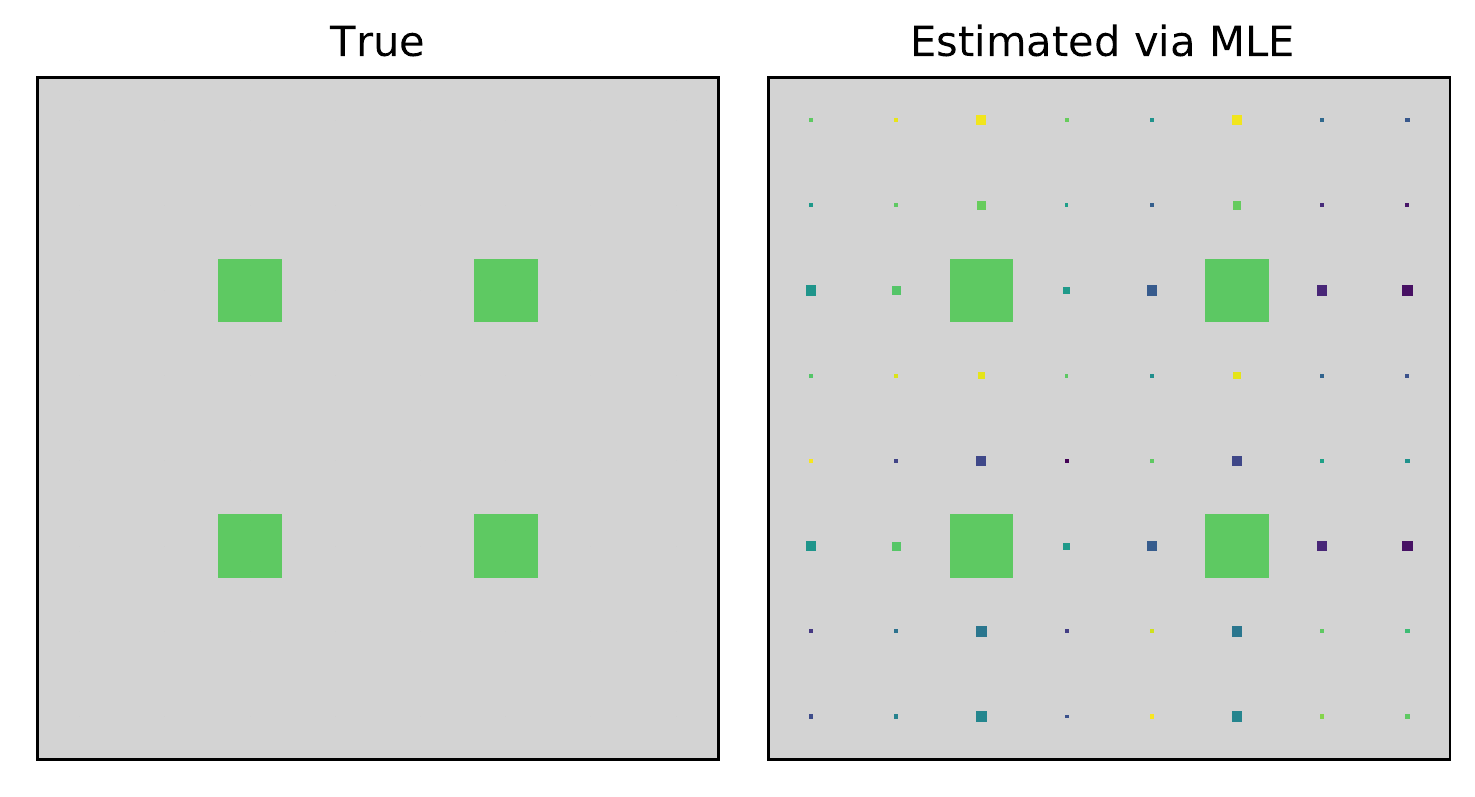}
    \caption{Comparison of the two dimensional projection plot between the estimated state $P_{453}$ and the true state.}
    \label{fX}
\end{figure}

\section{Gravitational Waves as a context}\label{sec4}
So far, we have described an alternative method to explain the entanglement paradox in the binary black hole system. We have incorporated the pseudo-density matrix formalism to explain this phenomenon. We have considered that Hawking radiation, which is the cause for the phenomenology of the black hole evaporation, can be well established from the pseudo-density formalism point of view in such a binary system, in agreement with the conjecture presented by \citep{monogamy}. For the analysis, we have considered PDO in terms of the Pauli operators for the three-qubit system, where two of them fall into the binary black hole system and get entangled there. We have used a quantum optical set up to demonstrate these phenomena by simulations using a quantum virtual machine. The state tomographic reconstruction shows that the pseudo-density operator can appropriately describe the correlation that violates monogamy.

The first detection of gravitational waves in 2015 \citep{ligo,ligo2} has opened many new possibilities for us in understanding many fundamentals of physics and the Universe. Recent works \citep{grav,hr}, have shown signs that there is a  scope for using the gravitational waves as an effective tool for understanding the Hawking radiation and probe into the black hole physics. Lately, works also show \citep{r1,r2,r3,r4,r5,r6} that it is very much possible to extend the standard framework of Hawking radiation in a single black hole to that of a binary black hole system (both non-spinning and spinning). In this context, upcoming gravitational waves detection programs like LISA \citep{lisa} are well designed. They will target objects like binary black hole systems typically a supermassive galactic black hole orbited by a stellar black hole \citep{lisa}. For such large mass ratio systems, \citep{hr} has shown how there will be Hawking radiation exchange between the two merging black holes and also that this exchange will not be attenuated by other physical parameters like the tidal force, relative motion, etc. In addition, they proposed that such exchanged Hawking radiation will lead to the production of gravitational waveforms different than those predicted by the classical theory of gravity and in future tests of gravitational waves, it is highly likely that such precision measurement can be recorded.

 Also in their paper, \citep{grav} has shown how binary gravitational systems can be expected to produce entangled signal emissions and how LIGO-like detectors can be used to detect them. In what follows, we put forward a thought experiment, trying to bridge this gap and make more use between the theoretical conjectures and the observational artefacts available from gravitational waves. We also explore it's verification possibilities.

In the work~\citep{hr}, the authors stated that owing to the effects of Hawking radiation from the binary black hole systems, the emitted gravitational waves will have a deviation in their characteristics from that predicted by the semi-classical theories of gravity. However, we suggest that the exchanged Hawking radiation between the two black holes will not hinder the normal entanglement process to propagate, exactly as outlined in our current work. We make an assumption that in an unlikely situation if simultaneously Hawking radiation and gravitational waves were both emitted from the outside neighborhood of the horizon of a binary black hole system, the entanglement information that would be imprinted in both these carriers would be the same, essentially describing the previous quantum state's information within the common envelope of the binary black hole's horizon \footnote{Although realistically that will never be possible to observe them simultaneously, as measurable Hawking radiation will be produced at a much much later stage in the lifetime of a black hole.}. A verification of this thought experiment is proposed with our optomechanical setup. If future observations of the gravitational waves are available with better precision, then we can replace the laser beam source in our optical setup with the characteristics waves of the gravitational waves (treating both as standard electromagnetic waves) and perform the optical simulations with the real data. In spite of the fact, that the gravitational waves detected are not the Hawking radiation waves from the black hole, but in the situation described above, they should carry the same entanglement imprint to that of Hawking radiation if they were simultaneous at the time of emission. If this is experimentally verified as we suggest with our optical setup, then we can do away with the requirement of detecting Hawking radiation separately for retrieval of quantum state information from inside the black hole. If the results provide satisfactory verification of the conjecture we proposed with good agreement between the theoretical and the experimental values, in our optical setup using the gravitational waves, then we will verify our above assumption.

This could potentially open new possibilities with the use of the gravitational waves as a tool for understanding the black hole paradox and information retrieval.  We can explore the possibility of understanding the quantum states of the particles inside the black hole which would be in spatial entanglement with the particles from outside (which in our case is particles $1$, $3$ and $4$) or in other words we can have the possibility to access the information of the inside of a black hole. The other possibility being, the gravitational waves detected being originated from the binary black hole system as explained before, if, via reverse engineering, the entanglement information which these waves will carry can be successfully segregated \citep{ent0,ent1}, we can also do a verification of our proposed conjecture and try to explore the same set of questions with a stronger benchmark.

\section{Conclusions \& Discussions}\label{sec5}
To conclude, we provided verification of the conjecture presented by \citep{monogamy} with a different system than theirs. We also tried to explore the possibilities of how their novel work could be brought to more practical setups, from where we can try to exploit our current available black hole observational information in the form of the gravitational waves and make use of our conjecture for its experimental verification as well as explore the idea of real black hole entanglement related observational experiments in near future.  We would also like to mention, that we have analyzed the post-merger equilibrium state (ring down) of a reduced binary black hole system. We have seen that our set up can reproduce the results presented in~\citep{monogamy} of a single black hole system under such conditions. Additionally, we have encountered some interesting results from our analysis like the fluctuations in the imaginary plots (see section \ref{sec3}). As discussed already, the origin of these fluctuations is expected to be not just due to noise but due to some effect of the system. We plan to continue the investigation on the origins of these fluctuations and it’s consequences. 

Recent new developments in open quantum systems have drawn our attention to the possibility to extend the current project from this perspective. The dynamics of a system interacting with an environment can be analyzed in the framework of open quantum systems. One can express the thermalization phenomena of the Hawking radiation from a Schwarzschild or a de Sitter spacetime from an open quantum system framework~\cite{liu12,hu12,lomb12}. Our model conjectured here can be suited to explore with open quantum systems.

\section*{Acknowledgements}
AM did this work with the grant from RK MES grant AP05135753, Kazakhstan. AM thanks Eric Linder for useful suggestions.  He also wants to thank the team from \href{https://rigetti.com/}{Rigetti Computing}    for helping him to run their quantum computer and also provide tutorials.



\end{document}